\documentclass[showpacs,twocolumn,pra,preprintnumbers ,amsmath, amssymb, superscriptaddress, aps]{revtex4-2}
\usepackage{color}
\usepackage{amsmath,amssymb}
\usepackage{pifont}
\usepackage{amssymb}  
\usepackage{bbold}
\usepackage{float}
\usepackage{subfloat}
\usepackage[caption=false]{subfig}
\usepackage{tikz}
\usepackage{makecell}
\usepackage{subfig}
\usepackage{pifont}   
\usepackage{graphicx} 
\graphicspath{{Figures2/}}
\usepackage{dcolumn}  
\usepackage{bm}       
\usepackage{multirow} 
\usepackage{placeins}
\usepackage[colorlinks]{hyperref}
\usepackage{mathtools}
\usepackage{appendix}

\def \be{\begin{align}}
	\def \ee{\end{align}}
\def \bea{\begin{eqnarray}}
	\def \eea{\end{eqnarray}}

\begin{document}
	\renewcommand{\thesection}{\arabic{section}}
	\renewcommand{\thesubsection}{\arabic{section}.\arabic{subsection}}
	\renewcommand{\thefigure}{\arabic{figure}}

	\title{
    Magnetic control of {Goos–Hänchen shifts and} group delay time in monolayer WSe$_2$}
	
	\author{Youssef Fattasse}
\affiliation{Laboratory of Theoretical Physics, Faculty of Sciences, Choua\"ib Doukkali University, PO Box 20, 24000 El Jadida, Morocco}
	\author{Rachid El Aitouni}
\affiliation{Laboratory of Theoretical Physics, Faculty of Sciences, Choua\"ib Doukkali University, PO Box 20, 24000 El Jadida, Morocco}
\author{Miloud Mekkaoui}
\affiliation{Laboratory of Theoretical Physics, Faculty of Sciences, Choua\"ib Doukkali University, PO Box 20, 24000 El Jadida, Morocco}
 \author{Pablo Díaz}
\affiliation{Departamento de Ciencias F\'{i}sicas, Universidad de La Frontera, Casilla 54-D, Temuco 4811230, Chile}  
\author{David Laroze}
\affiliation{Instituto de Alta Investigación, Universidad de Tarapacá, Casilla 7D, Arica, Chile}
\author{Ahmed Jellal}
\email{a.jellal@ucd.ac.ma}
\affiliation{Laboratory of Theoretical Physics, Faculty of Sciences, Choua\"ib Doukkali University, PO Box 20, 24000 El Jadida, Morocco}
	
	\begin{abstract}

We study the influence of an external magnetic field on the Goos–Hänchen (GH) shift and the group delay time (GDT) in monolayer WSe$_2$ in the presence of a magnetic barrier. The transport properties of Dirac-like carriers are obtained by solving the effective low-energy Hamiltonian and evaluating the corresponding transmission amplitudes. The GH shift and the GDT are subsequently extracted from the phase of the transmission coefficient. We systematically analyze their dependence on the magnetic field strength, incident energy, angle of incidence, and barrier width, with particular emphasis on the spin and valley degrees of freedom associated with the $K$ and $K'$ valleys. Our results show that the magnetic barrier strongly modulates both the GH shift and the GDT, leading to oscillatory behavior and pronounced spin–valley-dependent transport characteristics. Remarkably, the magnetic field enables selective control of the lateral shift and traversal time of carriers for each spin and valley channel, allowing for tunable spatial and temporal separation of electronic wave packets. This provides a mechanism for manipulating fermionic trajectories after transmission through the barrier in a highly controllable manner. Such tunability opens promising avenues for designing nanoscale devices based on spin and valley filtering, as well as for potential applications in information storage and processing within spintronic and valleytronic platforms.

\end{abstract}

	\maketitle
	
	\section{Introduction}	\label{Intro}

The research community currently views two-dimensional (2D) materials as an emerging field of scientific study. The ultra-thin structure of these materials sparks fundamental scientific interest due to their enhanced properties, which show potential for future applications in nanoelectronics, optoelectronics, and spintronics technologies \cite{ref1,ref2}. Graphene was the first material to attract great interest but its limited use in electronic applications is due to the absence of a bandgap in nature. Research started exploring other 2D materials, in particular transition metal dichalcogenides (TMDs), with the general formula MX$_2$ (M = Mo, W; X = S, Se, Te). These materials allow greater flexibility such as tunable band structures and strong spin-orbit coupling leading to a wider phenomena \cite{ref3,ref4}. Within this family, monolayer tungsten diselenide (WSe$_2$) has drawn special attention. Compared to graphene it has a direct bandgap that makes it particularly attractive for optical and electronic devices. Moreover, it has large spin splitting and the spin and valley degree of freedom are strongly interacting \cite{ref5,ref6}. Notable feature of WSe$_2$ is its intrinsic spin--valley locking. In other words, the electron spin is coupled with its valley state in such a way that charge, spin, and valley can be controlled together. This is quite the opposite of graphene, where such coupling is weak. Thus, WSe$_2$ exhibits various interesting quantum transport phenomena such as valley-dependent transport, spin polarization, and the valley Hall effect \cite{ref9,ref10}. For this reason, it is considered a promising material for devices beyond conventional electronics such as spintronics and valleytronics \cite{ref110,ref111}.

The study of charge transport in WSe$_2$ is therefore of fundamental importance for understanding its electronic behavior and optimizing device performance. In particular, external perturbations such as magnetic fields and potential barriers provide powerful tools to manipulate carrier dynamics \cite{ref11,ref12}. In analogy with graphene, magnetic barriers can induce quantum interference effects such as Fabry--Pérot resonances, resulting from multiple reflections of electron waves within the barrier region \cite{ref13,ref14}. Nonetheless, WSe$_2$ displays richer transport phenomena because of its finite bandgap and strong spin-orbit interaction, including spin- and valley-dependent tunneling and enhanced control of carrier polarization \cite{ref15,ref16}. Recently, the significance of Goos--Hänchen (GH) shifts and group delay time (GDT) in 2d materials has increased. Both GH shift and group delay time conveys rich information on the spacial and temporal aspects of quantum transport  \cite{ref17,ref18, ref161, ref162, ref163, ref164, ref165}. In systems with strong spin--valley coupling, such as WSe$_2$, these quantities become highly sensitive to both spin and valley indices, offering new possibilities for selective transport control \cite{ref22}.

The experimental investigation of the group tunneling time (GTT) in 2Dmaterials can be achieved through time-resolved and phase-sensitive measurements. Since the GTT corresponds to the Wigner--Smith phase delay associated with the transmitted wave packet, it can be extracted from the energy dependence of the transmission phase or from the temporal evolution of transmitted carriers. Ultrafast pump--probe spectroscopy and time-resolved optical techniques have been successfully employed to investigate carrier and exciton dynamics in transition-metal dichalcogenides, providing access to ultrafast temporal scales \cite{Chernikov2015,Wang2018}. Moreover, phase-sensitive transport measurements and interferometric approaches provide a way to probe quantum phase accumulation and delay times in mesoscopic systems \cite{Buttiker1983,Hauge1989}. Therefore, the GTT considered in this work represents an experimentally accessible quantity characterizing the temporal response of carriers crossing the WSe$_2$ barrier, rather than a classical transit time inside the barrier.

 {Recent studies have explored the control of Goos--H\"anchen (GH) shifts and group delay time (GDT) in 2D Dirac materials, particularly in graphene and silicene systems, where the effects of potential barriers, electric fields, and spin-dependent interactions have been investigated \cite{grap,sil}. However, the magnetic manipulation of GH shifts and GDT in transition-metal dichalcogenides remains less explored. In this work, we investigate the possibility of controlling GH shifts and group delay time in monolayer WSe$_2$ through a magnetic exchange field. The intrinsic spin--valley coupling of WSe$_2$ provides an additional degree of freedom for tuning carrier propagation, while the magnetic interaction offers an efficient external control parameter. Our approach goes beyond conventional transmission studies by demonstrating how magnetic effects can be employed to manipulate spatial and temporal beam shifts, opening new perspectives for spintronic and valleytronic applications.}

 We consider electron transport across a monolayer tungsten diselenide (WSe$_2$) under a magnetic barrier. For the system description, an effective continuum model for its electronic structure is adopted. We study the GH shift and the group delay time (GDT) concerning incident energy, magnetic field strength, incidence angle, and barrier width. Our results show that when electrons pass through the magnetic barrier, quantum interference can give rise to quasi-bound states giving rise to pronounced oscillatory behavior in both the GH shift and the group delay time. Spin–orbit coupling effects are crucial in WSe$_2$ \cite{ref23}. Additionally, the applied magnetic field breaks time-reversal symmetry, amplifying the asymmetry between the two inequivalent valleys ($K$ and $K'$) \cite{ref24, ref25, ref26, ref27, ref28}. The spin- and valley-transport characteristics become more pronounced. By varying external parameters such as the magnetic field or the barrier geometry, one is able to control the spatial and temporal evolution of charge carriers effectively. Finally, the aforementioned results contribute to identifying magnetic barriers as a robust tuning knob for quantum transport in 2D materials with potential impact on spintronic and valleytronic devices.

The paper is organized as follows. In Sec.~\ref{Theory}, we present the theoretical framework based on a continuum model to describe the electronic properties of monolayer WSe$_2$. We derive the energy spectrum in the three regions of the system under the influence of a magnetic barrier. In Sec. \ref{Grdelay}, we determine the transmission and reflection coefficients by applying the continuity conditions to the eigenspinors at the interfaces $x = 0$ and $x = L$. In Sec.~\ref{Grdelay}, we compute the Goos--Hänchen (GH) shift and the group delay time (GDT), and discuss their dependence on the relevant physical parameters for both valleys ($K$ and $K'$) and spin orientations. In Sec.~\ref{Num}, we present and analyze the numerical results for the GH shift and the group delay time, highlighting their sensitivity to the magnetic field, the incident energy, the incident angle, and the barrier width. In Sec.~\ref{GTNG}, we present a comparative analysis between monolayer WSe$_2$ and graphene, discussing the similarities and fundamental differences in their transport properties. Particular emphasis is placed on the role of spin--valley coupling in WSe$_2$, which is absent in graphene, and its impact on the Goos--Hänchen shift and the group delay time. Finally, Sec.~\ref{Concl} summarizes the main findings of and presents the concluding remarks.

\section{Theoritical Model}	\label{Theory}

	\begin{figure}[ht!]
	\centering
	\includegraphics[scale=0.3]{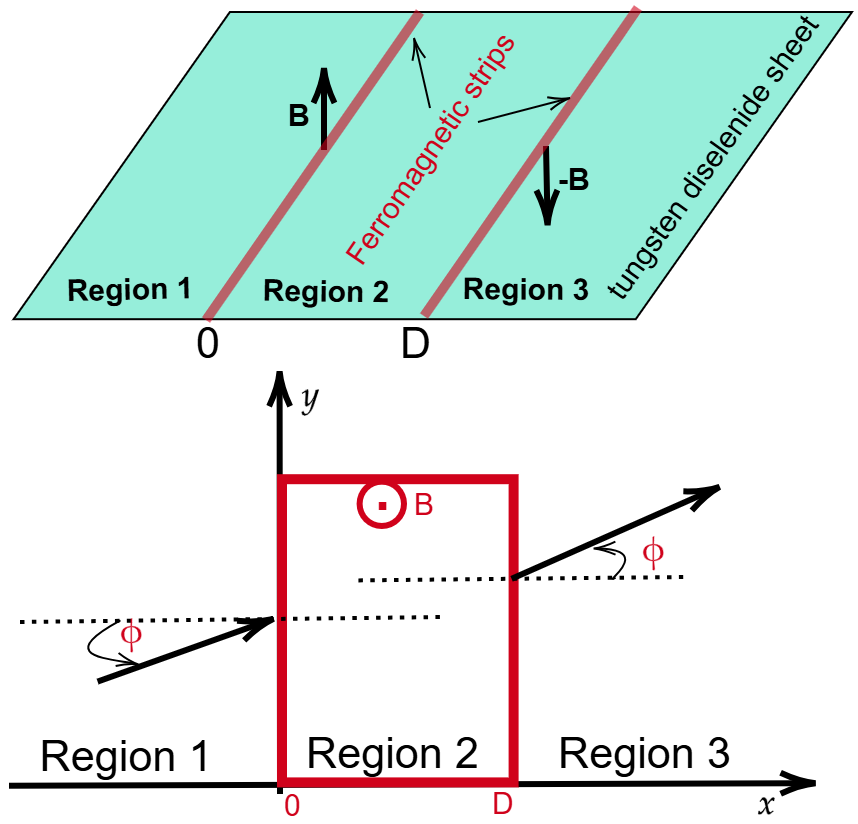}
	\caption{The schematic of two ferromagnetic strips separated by a distance $D$, deposited on a tungsten diselenide sheet. This configuration defines three distinct regions induced by the presence of the two strips.}\label{str}
\end{figure}
We investigate tunneling, group delay time, and the Goos–Hänchen spatial shift through a magnetic barrier in tungsten diselenide under a static magnetic potential. 
{Such a magnetic barrier can be experimentally realized by depositing two ferromagnetic strips on a monolayer WSe$_2$ sheet \cite{exp1,exp2}. The strips can be arranged symmetrically and placed sufficiently close to each other to generate a nearly uniform magnetic field region between them. This field acts as an effective magnetic barrier for carriers in WSe$_2$, modifying their spin- and valley-dependent transport properties. The corresponding group delay time can be accessed through the phase or time-resolved measurement techniques discussed above.
}
The barrier divides the WSe$_2$ sheet into three regions: regions 1 and 3 remain pristine, while the intermediate region is subjected to the magnetic field, as illustrated in Fig.~\ref{str}. In the presence of a magnetic field, an additional intrinsic perturbation arises: the Zeeman effect, which acts on both spin and valley degrees of freedom.

In the low-energy model, taking into account the Zeeman effect induced by the magnetic field, the intrinsic gap, as well as the intrinsic spin–orbit coupling, the Hamiltonian describing the motion of an electron through the proposed structure (Fig.~\ref{str}) is given by
\begin{widetext}
\begin{align}
	H =  v_F \left( \tau \sigma_x p_x + \sigma_y (p_y+eA_B \right) +\frac{\Delta}{2} \sigma_z + \lambda_c \tau s_z\frac{(\sigma_0+\sigma_z)}{2} +\lambda_v \tau s_z\frac{(\sigma_0-\sigma_z)}{2} +(s_zM_s-\tau M_v)\sigma_0
\end{align}
\end{widetext}
where $p_{x,y}$ are the momentum operators ($-i\hbar\frac{\partial}{\partial_{x,y}}$), $\sigma_i$($i=x,y,z$) are the Pauli matrices, $\tau=1(-1)$ for the valley $k$($k'$), and $s_z=1(-1)$ for spin up (down). Here, $v_F$ is the Fermi velocity ($v_F=5\times10^5$m/s), $\Delta=1.7$eV is the bandgap, $\lambda_{c,v}$ denote the spin–orbit coupling in both the conduction and valence bands  ($\lambda_v=112.5$ meV and $\lambda_c=7.5$ meV) \cite{WSe2valeurs}. $A_B$ is the vector potential component associated with the magnetic field created by two ferromagnetic strips, given by   $A_B=Bl_B\left[\Theta(x)-\Theta(x-D)\right]$, with $l_B=\sqrt{\frac{\hbar}{eB}}$ is the magnetic length and $\Theta(x)$ denotes the step function. Further,
$M_s$ and $M_v$ are the perturbations created by the magnetic field, 
known as the Zeeman effect \cite{Zemman}. They are defined as $M_{s,v} = \frac{g_{s,v} \, \mu_B \, B}{2}$, where $\mu_B = 5.788 \times 10^{-5} \, \text{eV/T}$ is the Bohr magneton, taking
 $g_s=2$ and $g_v=4$ \cite{gs,gv} for WSe$_2$.
For simplicity, we define  $E_c=\Delta/2+\lambda_{c} \tau s_z+s_zM_s-\tau M_v$, $E_v=-\Delta/2+\lambda_{v} \tau s_z+s_zM_s-\tau M_v$, $k_y^B=k_y+\frac{1}{l^2_B}$. 
We then express the Hamiltonian in matrix form
\small\begin{align}\label{H}
	H =
	\begin{pmatrix}
		E_c& v_F (\tau p_x - i\hbar k_y^B \\
		v_F \left(\tau p_x + i \hbar k_y^B\right) & E_v
	\end{pmatrix}.
\end{align}

Because the transverse wave vector $k_y$ is a conserved quantity, the spinor can be separated into longitudinal and transverse components. Accordingly, the spinor for the Hamiltonian $H$ can take the form $\Psi(x,y,t)=(\psi_c(x),\psi_v(x))^\dagger e^{ik_yy}$, denote the components of the wave function in conduction (c) and valence (v) bands, respectively. The result is thus reduced to effective  one-dimensional system along $x$-direction where the transverse $y$-motion is effectively subsumed in the conserved momentum $k_y$. 
 Hence, the resultant Hamiltonian can be formulated in terms of $k_y$, which in turn enables us to find the relevant eigenvalues. They are given by
\begin{align} \label{energie}
E^\pm = \frac{E_c + E_v}{2} \pm \sqrt{\left(\frac{E_c - E_v}{2}\right)^2 + v_F^2 \hbar^2 \left[ k_x^2 + ( k_y^B)^2\right]}
\end{align}
The resolution of the eigenvalue equations $H \Psi = E \Psi$ in the three distinct regions of the structure allows us to determine the explicit form of the spinor wave functions in each region. In particular, the system is divided into three regions: regions 1 and 3 correspond to pristine WSe$_2$, while region 2 represents the magnetic barrier. In regions 1 and 3, where no external magnetic perturbation is present, the electronic properties are governed solely by the intrinsic Hamiltonian of monolayer WSe$_2$. Consequently, the eigenvalue equations in these regions can be written in the following form
\begin{align}
	iv_F\hbar\left[ \tau\frac{\partial}{\partial x} - ik_y \right]\phi_v(x) &= \varepsilon_c  \phi_c(x) \\
	iv_F\hbar\left[\tau\frac{\partial}{\partial x} + k_y  \right]\phi_c(x) &=\varepsilon_v\phi_v(x)
\end{align}
with $\varepsilon_{c,v}=E-E_{c,v}$. 
From these equations, we obtain a second-order differential equation, whose solutions in regions 1 and 3 can be expressed in terms of propagating plane-wave states. These solutions yield the general forms
\begin{align}
    &\psi^1_{c,v}(x)=\left[\begin{pmatrix}
			1\\ e^{i\phi}
		\end{pmatrix}e^{i k_xx}+r\begin{pmatrix}
		1\\ -e^{-i\phi}
		\end{pmatrix}e^{-ik_xx}\right] \label{6}
\\
        &\psi^3_{c,v}(x)=t\begin{pmatrix}
			1\\ e^{i\phi}
		\end{pmatrix}e^{i k_xx} \label{7}
	\end{align}
where $r$ and $t$ denote the reflection and transmission coefficients, respectively. The wave vector component $k_x$ is defined as 
\begin{align}
    k_x =\tau \sqrt{\varepsilon_c \varepsilon_v/(\hbar v_F)^2-k_y^2}.
\end{align}
while the incident angle is $\phi=\tan^{-1}\left(\frac{k_y}{k_x}\right)$.

In region 2, where the magnetic field is applied, the eigenvalue equations retain the same form, except for the substitution $k_y \rightarrow k_y^B$. In a similar way, the two spinors of the wave function can be readily obtained, with a modified wave vector 
\begin{align}
    q_x =\tau \sqrt{\varepsilon_c \varepsilon_v/(\hbar v_F)^2-(k_y^B)^2}
\end{align}
The corresponding eigenspinor is given by
\begin{align}
        &\psi^2_{c,v}(x)=\left[a_2\begin{pmatrix}
			1\\ e^{i\theta}
		\end{pmatrix}e^{i q_xx}+b_2\begin{pmatrix}
		1\\ -e^{-i\theta}
		\end{pmatrix}e^{-iq_xx}\right]
	\end{align}
with the angle $\theta=\tan^{-1}\left(\frac{k_y^B}{q_x}\right)$.
In the following, we employ the derived energy spectrum to analyze the Goos--Hänchen (GH) shifts and the group delay time (GDT).

\section{GH shifts and GDT}\label{Grdelay}
The two ferromagnetic strips are considered infinite along the $y$-direction, that is, their length is assumed to be much greater than the barrier width $D$, in order to eliminate edge effects \cite{infini1,infini2}. This assumption simplifies the model by reducing the boundary conditions to a single relevant dimension. Thus, by taking into account the continuity of the eigenspinors at the two interfaces of the barrier, ($\Psi^1(0,y,t)=\Psi^2(0,y,t)$ and $\Psi^2(D,y,t)=\Psi^3(D,y,t)$), one can determine the reflection and transmission coefficients. As a result, we obtain
\begin{align}
	&1+r=a+b\\
	&e^{i\phi}-re^{-i\phi}=ae^{i\theta}-be^{-i\theta}\\
	&te^{ik_xD}=ae^{i q_xD}+be^{-i q_xD}\\
	&t e^{i\phi} e^{i k_xD}=ae^{i\theta} e^{i q_x D}-be^{-i\theta} e^{-i q_x D}.
	\end{align}
    This set of equations can be solved straightforwardly to obtain the transmission and reflection coefficients as follows
    \begin{align}
    &t=\frac{e^{-i D k_x }
\cos\phi\cos\theta }{
\cos\phi\cos\theta\cos(D q_x ) - i \sin(Dq_x ) (1-\sin\phi\sin\theta)
}\label{13}\\
&r=\frac{\sin(q_x D)(\sin\phi-\sin\theta)e^{i\phi}}{
	\cos\phi\cos\theta\cos(D q_x ) - i \sin(Dq_x ) (1-\sin\phi\sin\theta)}. \label{14}
\end{align}
It is convenient to write 
\begin{align}
    t=|t| e^{-\varphi_t}, \quad r= |r| e^{-\varphi_r}.
\end{align}
The continuity equation $\partial_t \rho + \nabla \cdot \mathbf{J} = 0$ allows us to determine the transmission and reflection probabilities because it proves that probability remains unchanged. The probability density is represented by $\rho = |\Psi|^2$ while the current density is represented by $\mathbf{J}$.
The relation enables us to express the incident and reflected and transmitted currents through the equations $J_{\text{in}} = 2 v_F \cos(\phi)$, $J_{\text{re}} = 2 v_F \cos(\phi), |r|^2$ and $J_{\text{tr}} = 2 v_F \cos(\phi), |t|^2$. The reflected and transmitted currents depend directly on the squared amplitudes of the reflection and transmission coefficients according to this statement. The reflection and transmission probabilities receive their definition through the formulas $R = \frac{J_{\text{re}}}{J_{\text{in}}}$ and $T = \frac{J_{\text{tr}}}{J_{\text{in}}}$,  which naturally simplify to $R = |r|^2$ and $T = |t|^2$, 
providing 
 \begin{align}
	&T=\frac{
		\cos^2\phi\cos^2\theta}{
		\cos^2\phi\cos^2\theta\cos^2(D q_x ) +\sin^2(D q_x ) (1-\sin\phi\sin\theta)^2
	} \label{15}\\
    &R = \frac{\sin^2(D q_x)\left[(1-\sin\phi \sin\theta)^2 - \cos^2\phi \cos^2\theta \right]}{
\cos^2\phi \cos^2\theta \cos^2(D q_x) + \sin^2(D q_x)(1-\sin\phi \sin\theta)^2} \label{16}.
\end{align}
The condition $R + T = 1$ shows that all incident particles in the system must either be reflected or transmitted without any system loss.

The Goos--Hänchen (GH) shifts and the group delay time provide valuable insight into the spatial and temporal transport properties of Dirac-like carriers in monolayer WSe$_2$. The GH shift, originating from the phase of the transmission coefficient, appears as a lateral displacement of the transmitted wave packet and exhibits oscillatory behavior with respect to the incidence angle, barrier width, and carrier energy due to quantum interference and quasi-bound states inside the potential barrier. The group delay time shows a similar resonant behavior, increasing near transmission resonances where carriers spend a longer time within the barrier region. While both quantities generally increase with the incident energy, incidence angle, and barrier width, higher barrier heights tend to suppress the delay time by reducing the transmission probability. These results demonstrate that electrostatic barriers offer an efficient way to control both the spatial and temporal dynamics of Dirac-like quasiparticles in WSe$_2$, highlighting their potential for applications in nanoelectronics, spintronics, and valleytronics.

The transmission (\ref{13}) and reflection (\ref{14}) coefficients can be expressed as
\begin{align}\label{Cnum}
	t=\sqrt{T} e^{i\varphi_{t}}, \quad
	r=\sqrt{R}e^{i\varphi_{r}}
\end{align}
where  $T$ and $R$ are given in (\ref{15}-\ref{16}), while 
the corresponding  phase shifts are
\begin{align}
	\varphi_{t}=\arctan\left(i\frac{t^{\ast}-t}{t+t^{\ast}}\right), \quad \varphi_{r}=\arctan\left(i\frac{r^{\ast}-r}{r+r^{\ast}}\right).	
\end{align}
	In what follows, we discuss the theoretical framework used to investigate the group propagation time in both transmission and reflection. In particular, a spatiotemporal wave packet can be employed to model a finite-duration electron beam as a weighted superposition of plane-wave spinors. Following \cite{chen08}, the incident, reflected, and transmitted beam waves at the interfaces 
    $ (x = 0, x = D) $ can be expressed as double Fourier integrals over the frequency $\omega$ and the transverse wave vector $k_y$. Representative expressions for these wave packets are given by
	\begin{align}
		&\label{int1}
		\Phi_{\sf in}(x,y, t)=\iint f(k_y,\omega)\ \Psi_{\sf in} (x,y) \  e^{-i\omega
			t}\ dk_yd\omega\\
		&\label{int2}
		\Phi_{\sf re}(x,y,t)=\iint  rf(k_y,\omega)\ \Psi_{\sf re} (x,y) \ e^{-i \omega
			t}\ dk_yd\omega
		\\
		&\label{int3}
		\Phi_{\sf tr}(x,y,t)=\iint  tf(k_y,\omega) \ \Psi_{\sf tr} (x,y) \ e^{-i \omega
			t}\ dk_yd\omega
	\end{align}
	in which the spinors $ \Psi_{\sf in}, \Psi_{\sf re} $ and  $ \Psi_{\sf tr} $ are  provided in \eqref{6}
	and \eqref{7}. According to \cite{Beenakker}, the angular spectral distribution is considered to have a Gaussian form $f(k_y,\omega)=w_ye^{-w_{y}^2(k_y-\omega)^2}$ with the half beam width at the waist being $\omega_y$, and the wave frequency is $\omega=E/\hbar$. Injecting \eqref{Cnum} into (\ref{int2}-\ref{int3}) to determine the total phases of the reflected and transmitted wave functions at ($x = 0 $, $x = D $). This process yields to 
	\begin{equation}
		\mathbf{\Phi}_{r}=\varphi_{r}+k_yy-\omega t, \quad \mathbf{\Phi}_{t}=\varphi_{t}+k_yy-\omega t.
	\end{equation}
	Using the stationary phase approximation \cite{Steinberg11, Li111}, i.e.  $  \frac{\partial\mathbf{\Phi}_{\gamma}}{\partial\phi}=0 $ and $  \frac{  \partial\mathbf{\Phi}_{\gamma}}{\partial\omega}=0 $, we calculate the GH shifts $ S_{\gamma} $  and  group delay time $ \tau_{\gamma} $.
	\begin{align}
		\label{ghss}	S_{\gamma}&=- \frac{\partial \varphi_{\gamma}}{\partial
			k_{y}}\\	
		     \tau_{\gamma}&= \tau^{\varphi_{\gamma}} +
		\tau^{s_{\gamma}}=\frac{\partial \varphi_{\gamma}}{\partial
			\omega}+\left(\frac{\partial k_y}{\partial
			\omega}\right)S_{\gamma}
	\end{align}
	where $\lambda=t, r$ stands for the transmission and reflection amplitudes, such that
	$ \tau^{\varphi_{\gamma}}$ denotes the time derivative of
	phase shifts, while  $\tau^{s_{\gamma}}$  denotes the contribution of  $S_{\gamma}$. Note that 
	$\tau_{\gamma}$  can be thought of as the average of the group delay times of the two components because the wave function involves a two-component spinor. As a result, we have 
	\begin{eqnarray}
		\tau^{\varphi_{t}}=\hbar \frac{\partial \varphi_t}{\partial
			E}+\frac{\hbar}{2}\frac{\partial \phi'}{\partial E}, \quad
		\tau^{\varphi_{r}}=\hbar \frac{\partial \varphi_r}{\partial E}
	\end{eqnarray}
	and in GH shifts
	\begin{equation}
		\tau^{s_{t}}=\frac{\sin\phi}{\upsilon_F}S_t, \quad
		\tau^{s_{r}}=\frac{\sin\phi}{\upsilon_F}S_r.
	\end{equation}
Next, we will numerically  present and analyze the GH shifts $S_t$ and group delay time $\tau_{t}/\tau_0$ for Dirac fermions in tungsten diselenide when scattered by a potential barrier based on the results obtained previously. For this purpose, we introduce the scaled Fermi wavelength
$\lambda = \frac{2\pi}{k_F}$ and the characteristic time scale $\tau_0 = \frac{d }{v_F}$. The quantity $\tau_0$ represents the time required for a free electron (without quantum effects) to traverse the barrier.

	It is worth noting that the group tunneling time is not a universal constant but depends on the incident energy, incidence angle, magnetic field, and barrier parameters. The values obtained in our calculations are expressed in the natural time scale of the system, $\tau_0=\hbar/E_0$. For the energy scales considered in monolayer WSe$_2$, this corresponds to delay times typically in the femtosecond regime, which is consistent with the characteristic timescale of ultrafast carrier dynamics in two-dimensional materials. The possibility of controlling this temporal delay through magnetic spin- and valley-dependent effects suggests potential applications in valleytronic and spintronic devices, where tunable carrier propagation, temporal filtering, and information processing can be achieved.

\section{NUMERICAL RESULT}\label{Num}

After obtaining the analytical expressions for the transmission amplitude, as well as the corresponding Goos–Hänchen shift and group delay time, we proceed to a detailed numerical analysis to explore their physical behavior under realistic conditions. By evaluating these quantities for a range of parameters, including the magnetic field strength, incident energy, angle of incidence, and barrier width, we illustrate how the magnetic barrier influences the transport properties of Dirac-like carriers in monolayer WSe$_2$. This numerical investigation allows us to visualize the oscillatory features of both the GH shift and the group delay time, and to highlight their strong dependence on spin and valley degrees of freedom associated with the $K$ and $K'$ valleys.

\begin{figure}[ht!]
	\centering
	\subfloat[{{$\phi=10^{\circ}$}}]{\centering\includegraphics[scale=0.23]{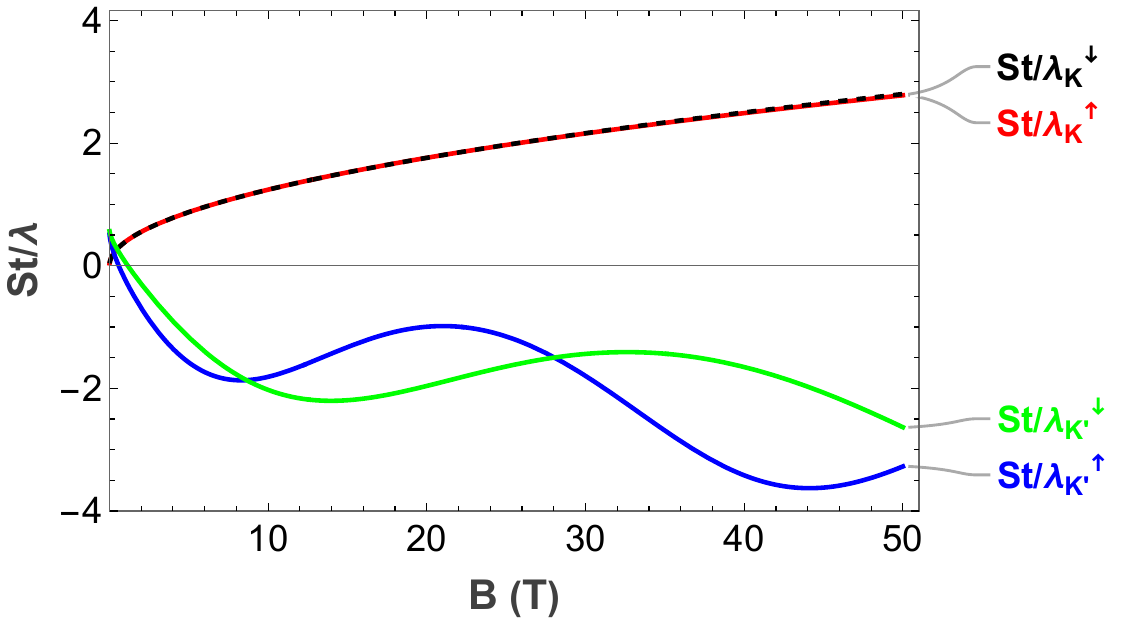}\label{fig2a}}
	\subfloat[{{$\phi=15^{\circ}$}}]{\centering\includegraphics[scale=0.23]{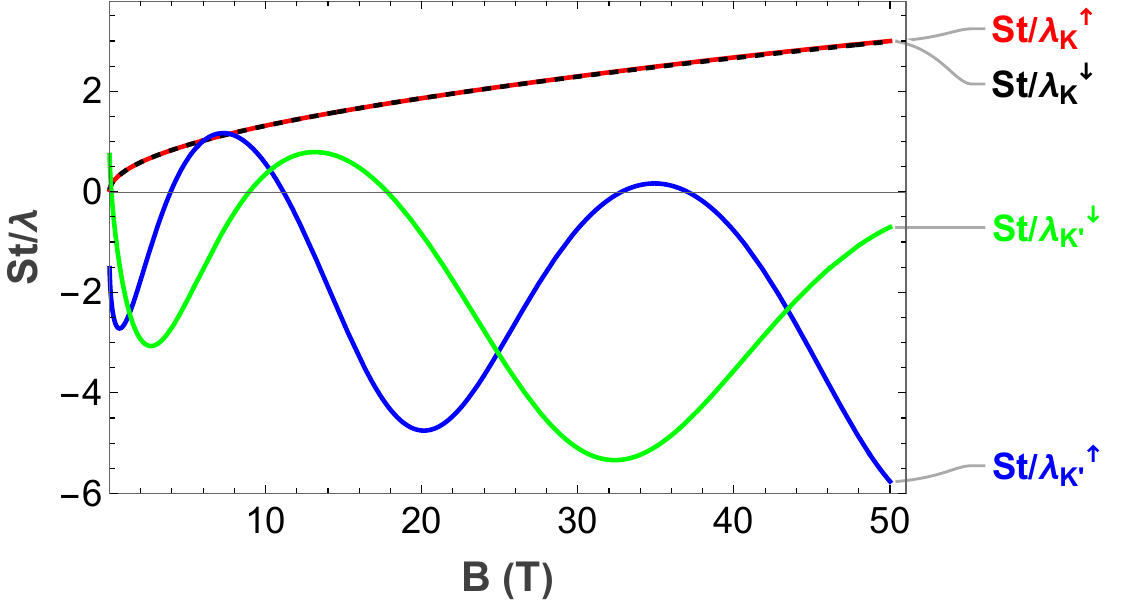}\label{fig2b}}\\
	\subfloat[{{$\phi=20^{\circ}$}}]{\centering\includegraphics[scale=0.23]{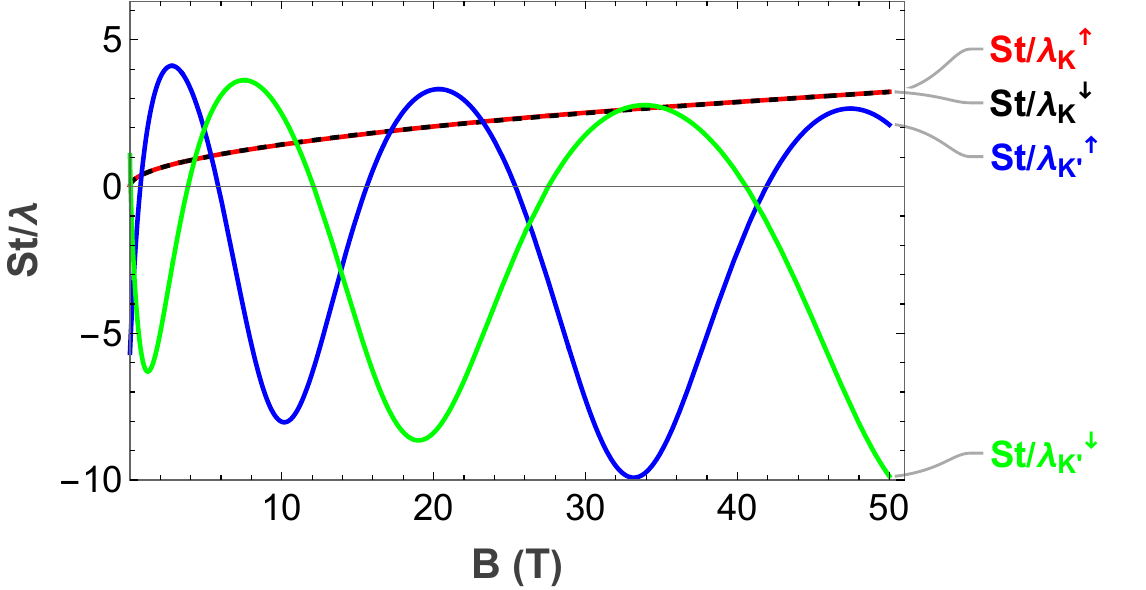}\label{fig2c}} 
	\subfloat[{{$\phi=30^{\circ}$}}]{\centering\includegraphics[scale=0.23]{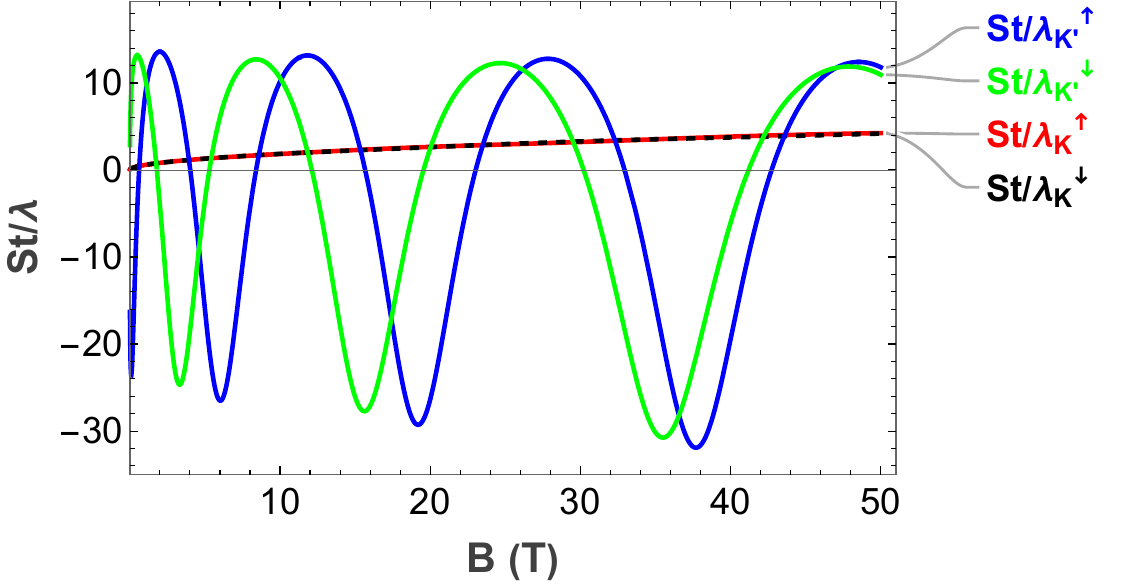}\label{fig2d}} 
	\caption{The GH shifts in transmissions  $S_t$/$\lambda$ versus the  magnetic field $B$ in the $K$ and $K'$ valleys for four values of  incident angles $\phi$, with incident energy {{$E=2.2$ eV}}, and barrier width {{$d=60$ nm}}.}
	\label{fig2}
\end{figure}

		Figure~\ref{fig2} presents the variation of the Goos--H\"anchen (GH) shifts with the magnetic field  $B$ for electron waves propagating through the 
		magnetic barrier in the $K$ and $K'$ valleys. The plots are performed for 
		$E=2.2\,\mathrm{eV}$ and $d=15\,\mathrm{nm}$, considering four different 
		incident angles, $\phi=10^\circ$, $15^\circ$, $20^\circ$, and $30^\circ$.
		At a small incident angle, $\phi=10^\circ$ in Fig.~\ref{fig2a}, the GH shifts in 
		the $K$ valley change smoothly with the magnetic field, where the spin-up and spin-down contributions remain almost identical. This indicates that the 
		magnetic barrier has only a weak influence on the spin degree of freedom in this valley. In contrast, the $K'$ valley shows a clear oscillatory response, with a 	small but visible separation between the two spin states. This different behavior between the two valleys comes from the valley-dependent phase acquired by the electron waves when crossing the magnetic region.
		For $\phi=15^\circ$ in Fig.~\ref{fig2b}, the GH shifts in the $K$ valley become smaller while maintaining its smooth behavior and weak spin dependence. On the other hand, the oscillations in the $K'$ valley become stronger, and the difference between the spin channels becomes more noticeable. This enhancement is related to the increase of the transverse wave-vector component $k_y$, which leads to stronger interference effects inside the magnetic barrier.
		When the incident angle is increased further to $\phi=20^\circ$ 
		in Fig.~\ref{fig2c}, the valley-dependent response becomes clearer. The $K$ 
		valley remains almost insensitive to the spin state, whereas the $K'$ valley 
		develops larger oscillations in the GH shifts. The increasing separation between the spin-up and spin-down states reflects the stronger influence of the magnetic field on the wave-packet dynamics.
		For the largest incident angle, $\phi=30^\circ$ in Fig.~\ref{fig2d}, the magnetic field produces the strongest modification of the GH shifts. In the $K'$ valley, the oscillations become more pronounced and the two spin contributions are clearly separated over a wide range of magnetic fields. This behavior results from the stronger interference effects associated with the larger transverse momentum. Meanwhile, the $K$ valley still exhibits a smooth and almost spin-independent response.
	These results highlight the important role of valley-dependent magnetic effects in controlling the GH shifts in monolayer WSe$_2$. The $K$ valley 
preserves nearly spin-degenerate behavior, while the $K'$ valley shows a strong 	spin-sensitive response. This contrast provides a route for manipulating the spatial displacement of electron wave packets through magnetic-field tuning.

\begin{figure}[ht!]
	\centering
	\subfloat[$E=1.3$~eV]{\centering\includegraphics[scale=0.225]{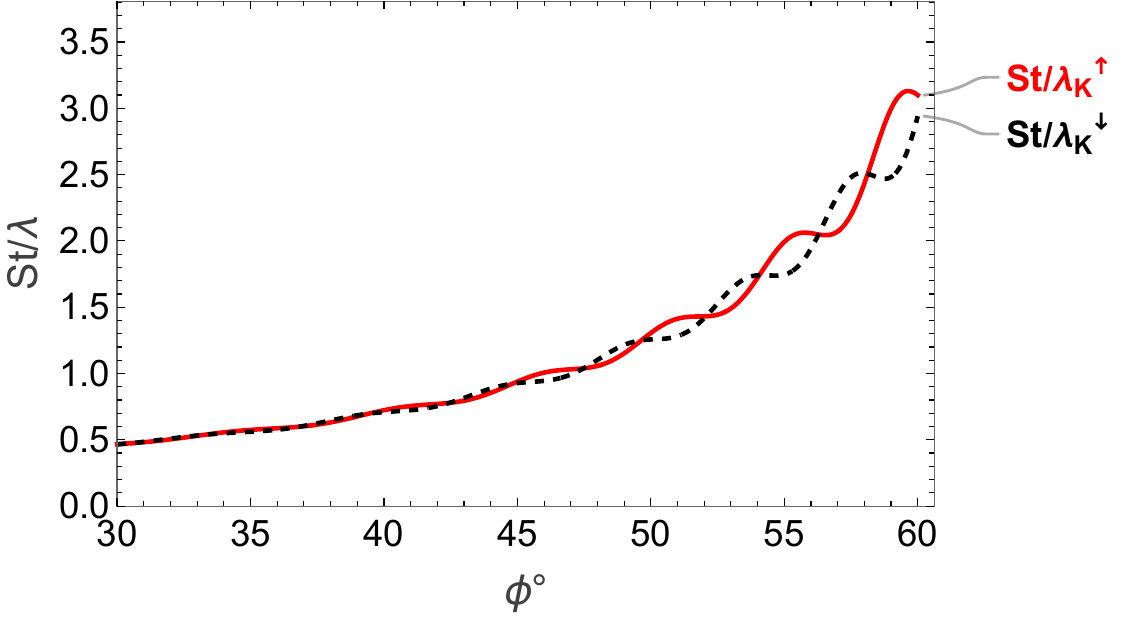}\label{fig3a}}
	\subfloat[$E=1.4$~eV]{
		\centering\includegraphics[scale=0.225]{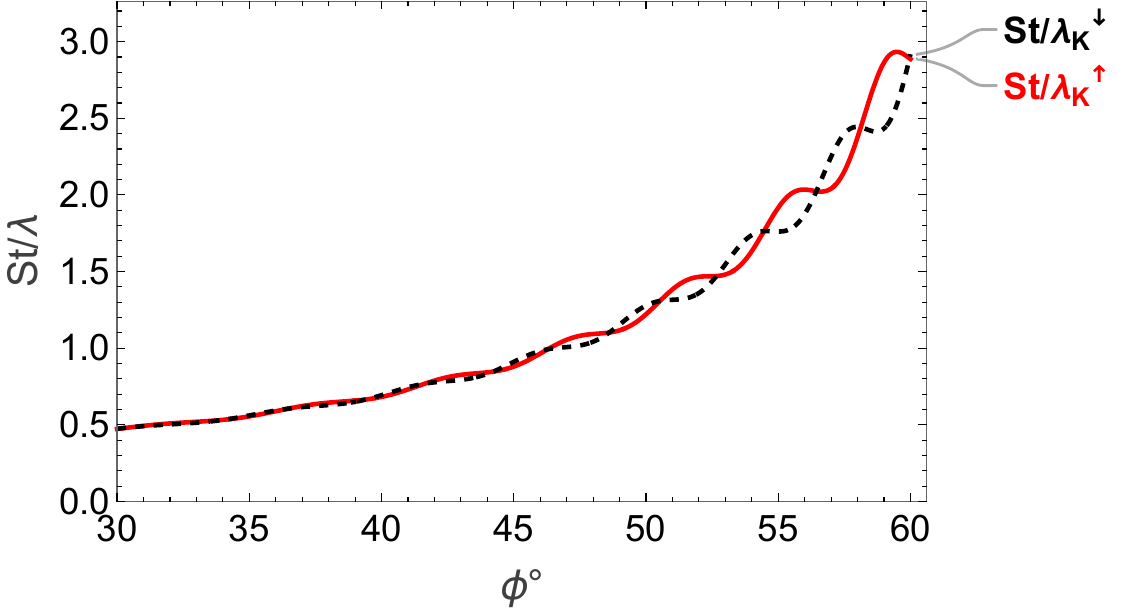}
		\label{fig3b}}  \\
	\subfloat[$E=1.3$~eV]{
		\centering\includegraphics[scale=0.225]{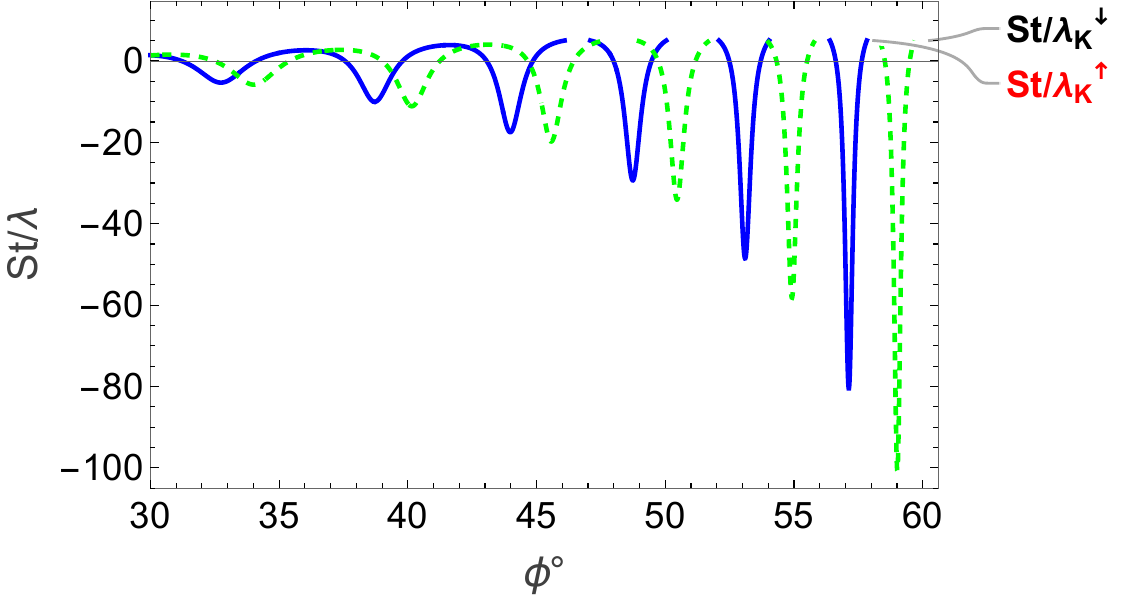}
		\label{fig3c}} 
	\subfloat[$E=1.4$~eV]{
		\centering\includegraphics[scale=0.225]{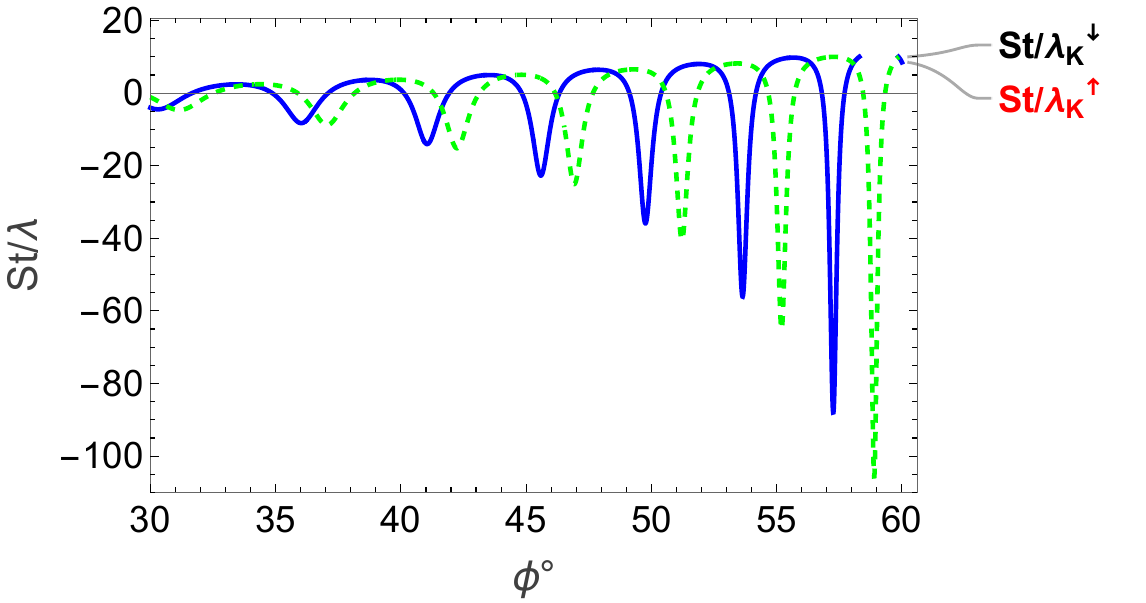}
		\label{fig3d}} 
	\caption{The GH shifts versus the incident angle for spin-up and spin-down electrons in the $K_{{\uparrow}{\downarrow}}$ and $K'_{{\uparrow}{\downarrow}}$ valleys at $E = 1.3$ and $1.4$ eV, with barrier width {{$d=15$ nm}}, and magnetic field {{$B=10 $ T}}.}
	\label{fig3}
\end{figure}

 Figure~\ref{fig3} shows what happens to the Goos–Hänchen (GH) shifts versus the incident angle for spin-up and spin-down electrons separately, in both the $K$ and $K'$ valleys, at energies of 1.3 and 1.4~eV 
 Figs.~(\ref{fig3a},\ref{fig3c}) and Figs.~(\ref{fig3b},\ref{fig3d}),
respectively, with barrier width {{15~nm}}, and  magnetic field {{B=10~T}}. It is found that the GH shift remains zero from normal incidence up to $\phi = 30^\circ$. For this reason, the plots are shown starting from $30^\circ$. 
The GH shifts vanish at normal incidence due to symmetry, and increase with the angle, reaching a maximum around {{$\phi \approx 60^\circ$}} as a result of enhanced interference within the barrier.
For larger incident angles, the GH shifts decrease rapidly as reflection becomes dominant and the transmission  is significantly suppressed.
The GH shifts also exhibit a clear dependence on the incident energy, since the electron energy directly modifies the wave vector, the transmission phase, and consequently the lateral displacement. A slight spin dependence is observed, particularly in the \(K\) valley, where the spin-up and spin-down components show small but noticeable differences in their peak positions and amplitudes. This behavior originates from the spin-dependent exchange interaction and spin-orbit coupling present in the system. In contrast, a pronounced valley asymmetry is clearly visible between the \(K\) and \(K'\) valleys. In the \(K\) valley, the GH shifts remain relatively small and positive Figs.~(\ref{fig3a},\ref{fig3b}), whereas in the \(K'\) valley they become significantly larger and negative Figs.~(\ref{fig3c},\ref{fig3d}). This opposite sign reflects the valley-dependent phase response of Dirac electrons under the combined influence of the magnetic barrier, exchange field, and intrinsic spin--orbit coupling. Such a strong valley contrast demonstrates the possibility of controlling both spin and valley degrees of freedom through the GH effect, which is of particular interest for potential applications in spin-valley filtering and valleytronic devices.

\begin{figure}[ht!]
	\centering
	\subfloat[$d=15$~nm]{\centering\includegraphics[scale=0.31]{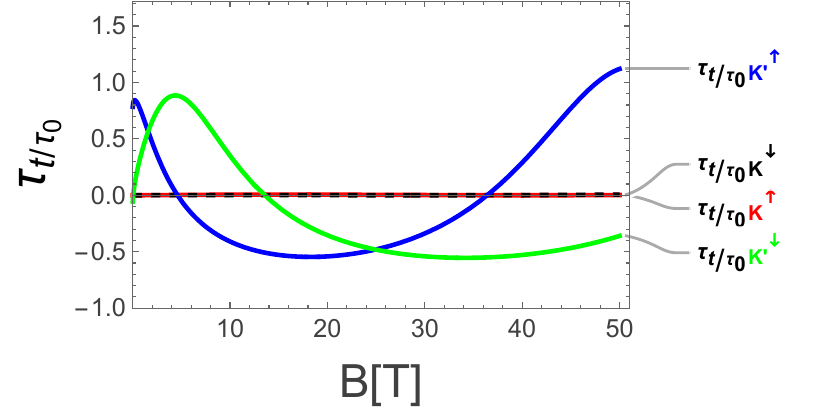}\label{fig4a}}
	\subfloat[$d=30$~nm]{
		\centering\includegraphics[scale=0.31]{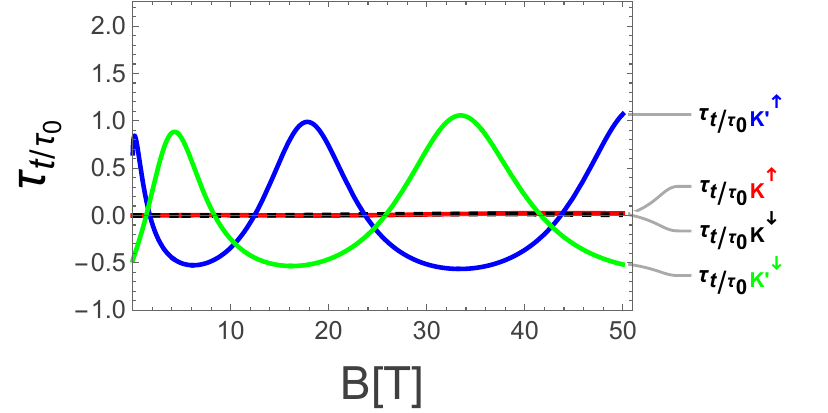}
		\label{fig4b}}  \\
	\subfloat[$d=45$~nm]{
		\centering\includegraphics[scale=0.31]{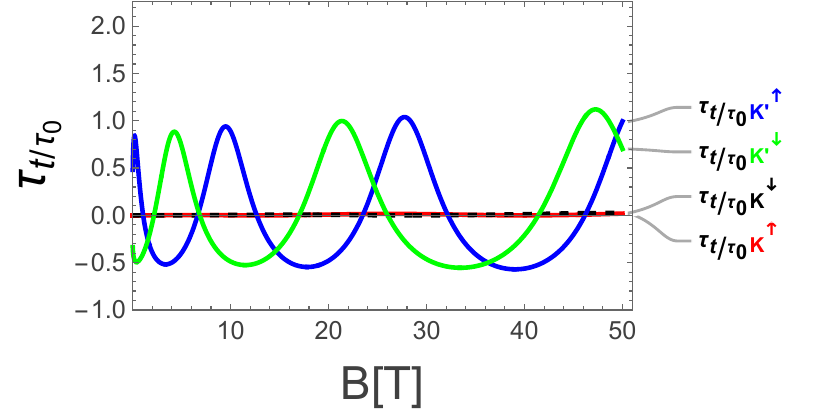}
		\label{fig4c}} 
	\subfloat[$d=60$~nm]{
		\centering\includegraphics[scale=0.31]{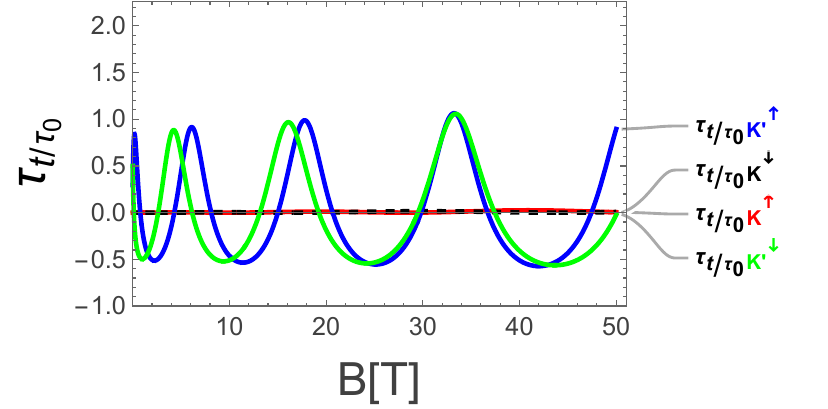}\label{fig4d}
	} 
	\caption{The group delay time in transmission $\tau_{t}/\tau_{0}$ versus the magnetic field $B$ in the $K$ and $K'$ for  incident energy $E=1.2$~eV,  incident angle $\phi=30^\circ$, and four values of the barrier width $d$. }
	\label{fig4}
\end{figure}

Figure~\ref{fig4} illustrates the group delay time (GDT) $\tau_t/\tau_0$ as a function of the magnetic field $B$ for different barrier widths $d = 15$, $30$, $45$, and $60$ nm shown in Figs.~(\ref{fig4a}, \ref{fig4b}, \ref{fig4c}, \ref{fig4d}), respectively, considering both valleys ($K$ and $K'$) and spin configurations (spin-up and spin-down). The results reveal a pronounced oscillatory behavior associated with Fabry--Pérot-type resonances induced by quantum interference inside the barrier region.
For the narrow barrier shown in Fig.~\ref{fig4a} with $d = 15$ nm, the GDT exhibits relatively few oscillations with moderate amplitudes. The resonant peaks are well separated, indicating weak confinement within the barrier. Moreover, the oscillatory behavior differs between the two valleys and spin orientations, reflecting the combined influence of valley-dependent and spin-dependent transport. The large spacing between successive peaks indicates a relatively large oscillation period in the magnetic field.
As the barrier width increases to $d = 30$ nm in Fig.~\ref{fig4b}, the number of resonant peaks increases significantly, while the oscillation period decreases. This behavior is attributed to stronger quantum confinement within the barrier region, which enhances multiple reflections of carriers. Consequently, the interference becomes more pronounced, leading to additional Fabry--Pérot resonances. Furthermore, the amplitudes of the oscillations become slightly larger, indicating enhanced dwell time of carriers inside the barrier.
For $d = 45$ nm in Fig.~\ref{fig4c}, the oscillatory behavior becomes even more pronounced. The number of resonant peaks increases further, and the oscillation period continues to decrease. This confirms that increasing the barrier width strengthens the quantum interference effects. Additionally, the maxima of the GDT become higher, suggesting that the carriers spend longer times within the barrier region before transmission, which is a typical signature of resonant tunneling processes.
Finally, for the largest barrier width $d = 60$ nm shown in Fig.~\ref{fig4d}, the oscillations become very dense with a substantial increase in the number of peaks. The period of oscillation becomes smaller, and the resonant amplitudes reach higher values. This behavior indicates strong confinement and enhanced Fabry--Pérot resonances inside the barrier. Moreover, the difference between the valleys and spin configurations remains noticeable, suggesting persistent valley- and spin-dependent transport characteristics under strong magnetic confinement.
Interestingly, it is observed that the GDT for the $K$ valley 
remains nearly zero for both spin-up and spin-down states, regardless of the 
barrier width or the applied magnetic field. This behavior suggests that the 
transport dynamics for the $K$ valley are weakly affected by the considered 
physical parameters, whereas the $K'$ valley exhibits pronounced oscillatory 
features. This result highlights a strong valley-dependent transport behavior, 
which may be useful for valleytronic applications.
In general, increasing the width of the barrier leads to a systematic increase in the number of resonant peaks, a greater increase in the amplitudes of the oscillation, and a reduction in the oscillation period. These features confirm that the barrier width plays a crucial role in controlling quantum interference effects and GDT in the presence of a magnetic field.

\begin{figure}[ht!]
	\centering
	\subfloat[$E=1.4$~eV]{
		\centering
		\includegraphics[scale=0.3]{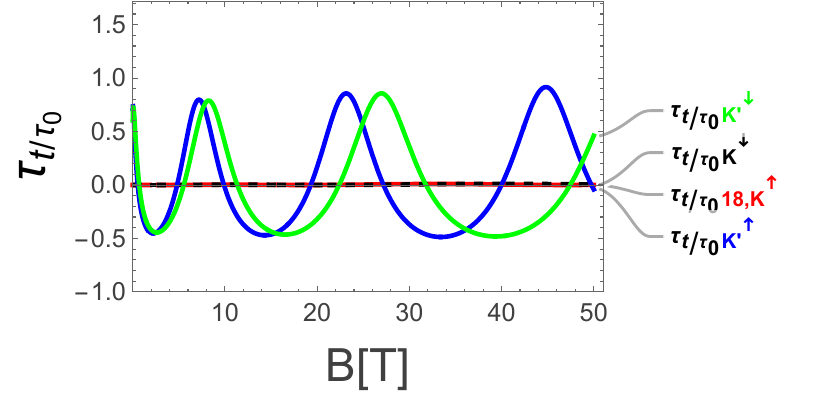}\label{fig5a}}
	\subfloat[$E=1.5$~eV]{
		\centering\includegraphics[scale=0.3]{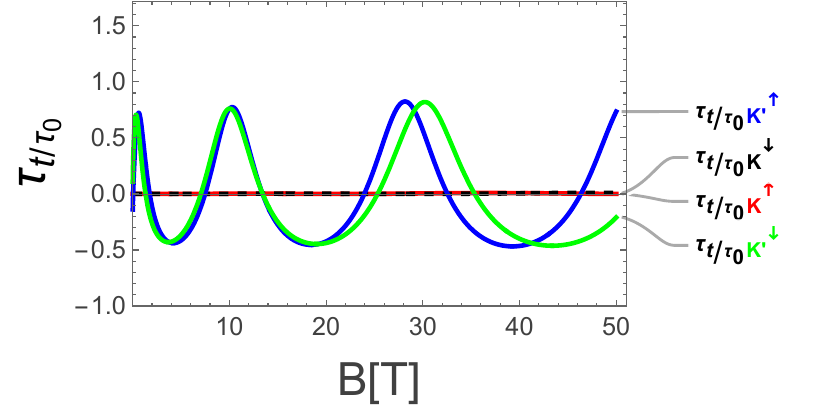}
		\label{fig5b}}  \\
	\subfloat[$E=1.6$~eV]{
		\centering\includegraphics[scale=0.3]{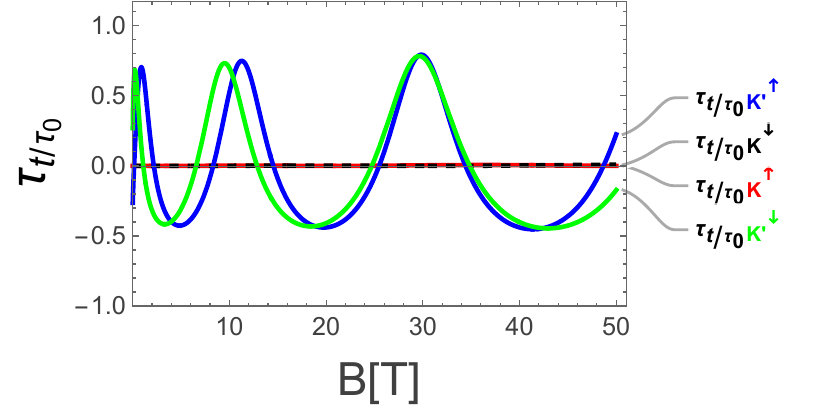}
		\label{fig5c}} 
	\subfloat[$E=1.8$~eV]{
		\centering\includegraphics[scale=0.3]{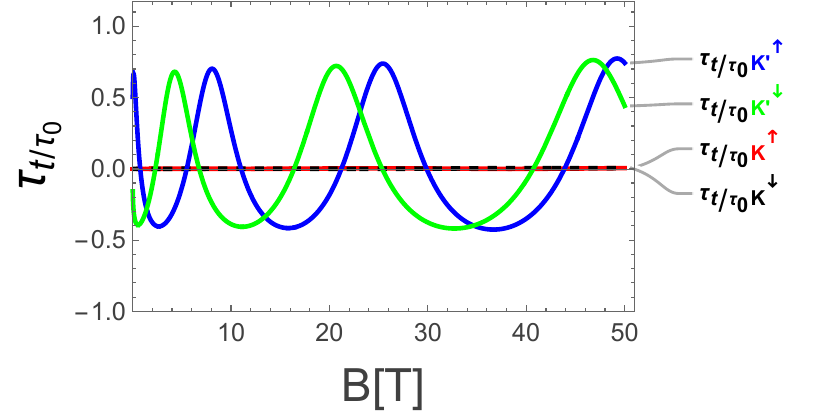}
		\label{fig5d}} 
	\caption{The group delay time in transmission $\tau_{t}/\tau_{0}$ versus
		the magnetic field $B$ in the $K$ and $K'$  for incident angle $\phi=30^\circ$, with different values of the 
		energy $E$ and barrier width $d=50$.}
	\label{fig5}
\end{figure}

Figure~\ref{fig5} displays the  group delay time (GDT) in transmission $\tau_t/\tau_0$ as a function of the magnetic field $B$ for fixed barrier width $d=50$~{nm} and incident angle $\phi=30^\circ$, and different incident energy $E={(1.4, 1.5, 1.6, 1.8)}\ \text{eV}$, as shown in Figs.~(\ref{fig5a}, \ref{fig5b}, \ref{fig5c}, \ref{fig5d}), respectively.
The results exhibit behavior similar to that observed previously, characterized by a sequence of resonant peaks as the magnetic field increases. 
These peaks originate from quasi-bound states formed inside the barrier, where constructive interference between multiple reflected electron waves enhances the transmission phase sensitivity. Since the group delay time is directly related to the energy derivative of the transmission phase, each resonant condition gives rise to a sharp increase in the GDT. 
 Away from resonance, the transmission probability decreases significantly and the transmission phase varies weakly with energy. Consequently, the GDT decreases and can approach zero in regions of the magnetic field where tunneling through the barrier is significantly inhibited. In particular, in the $K$ valley, the GDT nearly vanishes over certain intervals of $B$.  This behavior is associated with the appearance of evanescent modes inside the barrier, for which the longitudinal wave vector becomes imaginary. In such a regime, the propagating states are replaced by exponentially decaying solutions, so the phase accumulation through the barrier is strongly reduced and the corresponding delay time becomes very small.
The spin-up and spin-down channels display almost similar GDT profiles at comparatively weak magnetic fields. This is because the Zeeman splitting is still small in relation to the intrinsic spin-orbit interaction. Their resonant conditions match almost perfectly and their two spin-dependent dispersion branches remain almost degenerate as a result. But as the magnetic field grows, the Zeeman contribution gets more apparent and elevates this near-degeneracy. As a result, the resonance peaks for the two spin orientations divide and move differently, producing a distinct spin-dependent separation in the delay-time spectrum.
The number, position, and amplitude of the resonant peaks change sensitively with the incoming energy, which is yet another significant aspect. The propagating window inside the barrier is changed as $E$ rises, which alters the matching criteria at the interfaces and moves the quasi-bound-state energies. This causes the resonance pattern with magnetic field to be redistributed. Higher incident energies specifically let more propagating states inside the barrier, hence widening the scope of magnetic fields over which resonant transmission can occur.

\begin{figure}[ht!]
	\centering
	\subfloat[{$\phi=10^{\circ}$}]{
		\centering
		\includegraphics[scale=0.31]{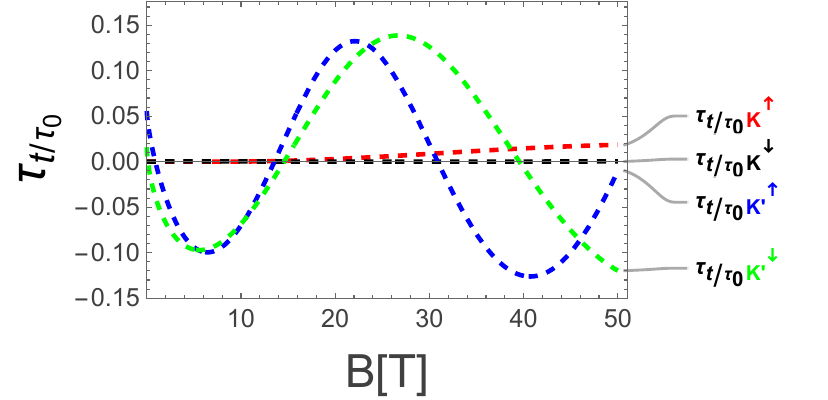}\label{fig6a}}
	\subfloat[{$\phi=15^{\circ}$}]{
		\centering\includegraphics[scale=0.31]{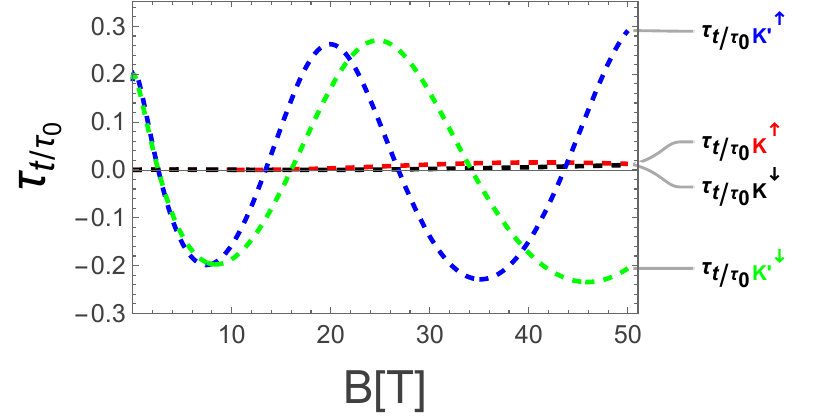}\label{fig6b}
	}  \\
	\subfloat[{$\phi=20^{\circ}$}]{
		\centering\includegraphics[scale=0.31]{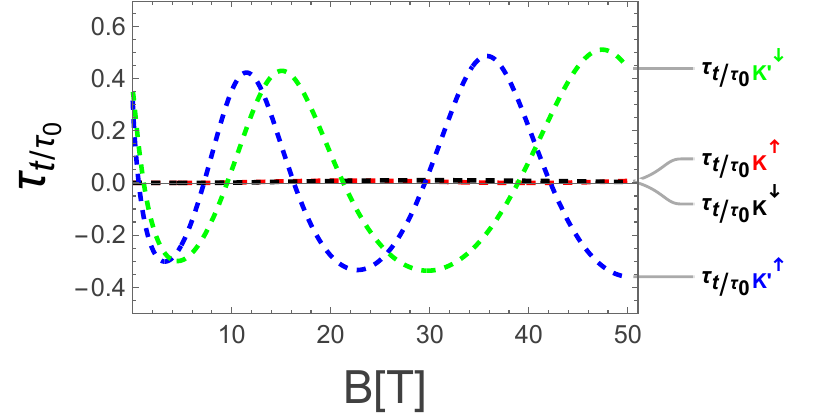}\label{fig6c}
	} 
	\subfloat[{$\phi=30^{\circ}$}]{
		\centering\includegraphics[scale=0.31]{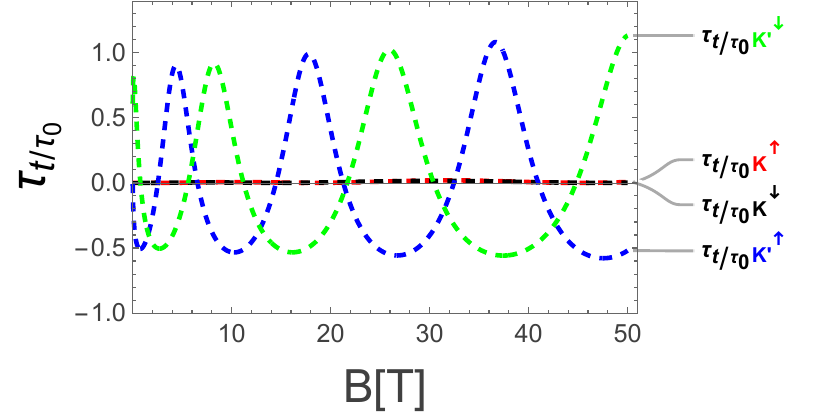}\label{fig6d}
	} 
	\caption{The  group delay time in transmission $\tau_{t}/\tau_{0}$ versus the magnetic field $B$ in the $K$ and $K'$ valleys for  incident energy $E=1.2$~eV,  barrier width $d=50$~nm, and different values of the incident angle $\phi$. }
	\label{fig6}
\end{figure}

{Figure~\ref{fig6} depicts  the group delay time (GDT) $\tau_t/\tau_0$ as a function of the magnetic field $B$ for $d=50\,\mathrm{nm}$ and $E=1.2\,\mathrm{eV}$. The results are presented for four incident angles, namely $\phi = 10^\circ$, $15^\circ$, $20^\circ$,
	For a small incident angle, $\phi=10^\circ$ in Fig.~\ref{fig6a}, the GDT exhibits weak oscillations with magnetic field in both valleys.  
	The small amplitude of the oscillations indicates that the magnetic barrier induces only moderate modifications of the transmission phase.
	A valley-dependent response is already visible, with the $K'$ valley showing larger delay times than the $K$ valley. The spin-up and spin-down contributions remain almost overlapped in the $K$ valley, while a small spin splitting appears in the $K'$ valley.
	When the incident angle increases to $\phi=15^\circ$ in Fig.~\ref{fig6b}, the oscillatory behavior becomes more pronounced. The increase of the transverse wave-vector component enhances the interference effects inside the magnetic barrier, leading to larger variations of the GDT, especially in the $K'$ valley. In contrast, the $K$ valley remains weakly affected, with only a small spin dependence.
	At $\phi=20^\circ$ in Fig.~\ref{fig6c}, 
	the magnetic-field dependence becomes stronger, with larger oscillation amplitudes and more pronounced extrema.
	This indicates a stronger modulation of the transmission phase and, consequently, of the GDT. The difference between the two valleys becomes more visible, and the spin splitting in the $K'$ valley is significantly enhanced.
	for a larger incident angle $\phi=30^\circ$ [Fig.~\ref{fig6}(d)], the 
	GDT exhibits its strongest oscillatory behavior. Large positive and negative 
	delay times appear over the investigated magnetic-field range, showing that the 
	carrier propagation can be strongly accelerated or delayed depending on the 
	magnetic-field strength. The pronounced separation between spin-up and spin-down 
	states in the $K'$ valley indicates an efficient spin-selective control of the 
	group delay time. In contrast, the $K$ valley remains comparatively stable with 
	much weaker oscillations and smaller spin dependence.}

\section{Tungsten diselenide vs graphene }\label{GTNG}

The study of Goos--H\"anchen (GH) shifts and group delay time (GDT) in two-dimensional Dirac materials has attracted significant interest because these quantities provide important information about the quantum transport behavior of charge carriers. The GH shift represents the lateral displacement of an electron beam after reflection or transmission through a potential or magnetic barrier, while the group delay time describes the temporal delay experienced by transmitted electrons due to phase accumulation inside the barrier region \cite{ref19,Song2012}. The two quantities show direct connection to transmission phase which helps in studying resonant tunneling and interference effects and spin- and valley-dependent transport phenomena.

Graphene and tungsten diselenide monolayer WSe$_2$, two two-dimensional materials, have quite different behaviors due to their different electronic structures. Charge carriers act as massless Dirac fermions with a linear energy dispersion close the Dirac points found at the \(K\) and \(K'\) valleys \cite{Novoselov2005, CastroNeto2009}. Because graphene has very little intrinsic spin-orbit interaction, the spin-up and spin-down states stay nearly degenerate \cite{Kane2005}.
For this reason, the GH shifts and the group delay time usually exhibit very similar behavior for both spin states in each valley. The valley degree of freedom can still strongly influence the transport process, especially in the presence of electrostatic or magnetic barriers, where the transmission phase becomes valley dependent and produces measurable valley-selective GH shifts \cite{Song2013,Wu2013}. However, because the spin splitting is very small, spin filtering in graphene remains relatively weak.

On the other hand, tungsten diselenide monolayer WSe$_2$, a member of the transition metal dichalcogenides (TMDs), displays great Dirac fermions and pronounced intrinsic spin--orbit coupling \cite{Xiao2012, Mak2014}. Unlike graphene, WSe$_2$ has a big direct band gap and robust spin splitting in both the valence and conduction bands. This great spin-orbit coupling causes spin-valley locking, which links the spin orientation straight to the valley index \cite{Xu2014}. Consequently, transport characteristics including the GH shifts and the group delay time become highly reliant on both spin and valley degrees of freedom. The spin-up and spin-down states are no longer equal in the \(K\) and \(K'\) valleys, hence there are significant differences in the strength, polarity, and resonance conditions of the GH shifts and GDT. This makes WSe$_2$ far more efficient for spin and valley filtering than graphene, therefore enabling more precise control of electron transport based on spin and valley indexes \cite{ref22,Zibouche2014}.

Due to its very high carrier mobility, high electrical conductivity, great thermal characteristics, and somewhat easy production method \cite{ref16}, graphene still ranks among the most commonly used substances in nanoelectronics from the perspective of the device. These properties make graphene especially suited for nanoscale sensors, flexible electronics, and high-speed transistors. Its poor intrinsic spin--orbit coupling restricts its effectiveness in spintronics applications unless more means like proximity effects, substrate engineering, or external magnetic fields are added \cite{Avsar2014}. WSe$_2$ on the other hand has greater promise for applications in spintronics and valleytronics since its inherent spin-valley coupling inherently permits effective control of spin and valley polarization without need for strong outside disturbances \cite{ Schaibley2016}. Furthermore appealing for low-power electronic devices, valley filters, and spin-selective transport systems is WSe$_2$'s large band gap and notable spin splitting.

Therefore, although graphene remains the most established and widely used material for conventional electronic applications, WSe$_2$ appears to be a more promising candidate for future spintronic and valleytronic technologies. Comparing GH shifts and group delay time in these two systems helps clarify the role of spin--orbit coupling and valley physics in quantum transport, and provides useful theoretical guidance for designing next-generation nanoscale devices based on controllable spin and valley degrees of freedom.

\section{Conclusion}\label{Concl}

We investigated the electronic transport properties of charge carriers in monolayer WSe$_2$ in the presence of a magnetic barrier. In particular, we analyzed two fundamental quantum transport quantities, namely the Goos–Hänchen (GH) shifts and the group delay time, as functions of the external magnetic field for different physical parameters such as the incident energy, the incident angle, and the barrier width. The analysis was carried out for both valleys ($K$ and $K'$) and for the two spin orientations (spin-up and spin-down), allowing us to explore the spin–valley-dependent transport properties of this two-dimensional material. 

First, we examined the behavior of the Goos–Hänchen shifts. Our results show that the GH shifts strongly depend on the magnetic field as well as on the incident parameters. Interestingly, we found that the GH shifts remain positive in the $K$ valley and do not allow a clear distinction between the two spin orientations. In contrast, the behavior in the $K'$ valley is markedly different. In this valley, the GH shifts become negative and exhibits a clear spin-dependent separation, which indicates the presence of a spin filtering mechanism induced by the magnetic barrier. This result highlights the possibility of achieving spin–valley selective transport in WSe$_2$, which is of particular interest for valleytronics and spintronic applications. 
Second, we analyzed the group delay time associated with the transmission process. The results reveal an oscillatory dependence of the normalized group delay time on the magnetic field, characterized by a series of resonant peaks. The number and amplitude of these peaks depend strongly on the incident energy, the incident angle, and the barrier width. From a quantum mechanical perspective, these oscillations originate from interference effects between electron wave components inside the magnetic barrier region. Moreover, the magnetic field modifies the electronic states through spin and valley Zeeman interactions, which leads to valley-dependent phase accumulation and consequently to distinct delay times for different spin–valley configurations. 

Overall, our findings demonstrate that the presence of a magnetic barrier can significantly modify both the GH shifts and the group delay time in monolayer WSe$_2$. In particular, the observed spin-dependent GH shifts in the $K'$ valley and the tunability of the group delay time by the magnetic field suggest that magnetic barriers provide an efficient mechanism for controlling spin and valley degrees of freedom in transition metal dichalcogenides. These results are consistent with previous studies on quantum transport and beam shifts in two-dimensional materials, where similar oscillatory behaviors and spin–valley effects have been reported. For example, the theoretical investigations of GH shifts in graphene and transition metal dichalcogenides have shown that external fields and barrier structures can strongly influence the lateral shift and transmission properties of Dirac fermions \cite{wu2009Conc} and \cite{chen2011Conc}. In addition, several works have demonstrated that magnetic barriers can induce spin- and valley-dependent transport phenomena in two-dimensional materials, opening new possibilities for designing spin–valley filters and quantum electronic devices \cite{Yes2016}. Therefore, the present study provides further insight into the quantum transport properties of WSe$_2$ under magnetic confinement and may contribute to the development of future nanoscale devices based on spin and valley control. 



\begin{thebibliography}{99}
\bibitem{ref1}F. Xia, H. Wang, D. Xiao, M. Dubey, and A. Ramasubramaniam, Nat. Photon. 8, 899 (2014).

\bibitem{ref2} Y. Liu, Y. Huang, and X. Duan, Nature 567, 323–333 (2019).
\bibitem{ref3} T. Chowdhury, E. C. Sadler, and T. J. Kempa, Chem. Rev. 120, 12563–12591 (2020).
\bibitem{ref4} 
Q. Fu, J. Han, X. Wang, P. Xu, T. Yao, J. Zhong, W. Zhong, S. Liu, T. Gao, Z. Zhang, L. Xu, B. Song  Adv. Mater. 33, 1907818 (2021).
\bibitem{ref5}R. Yang, Y. Fan, Y. Zhang, L. Mei, R. Zhu, J. Qin, J. Hu, Z. Chen, Y. H. Ng, D. Voiry, S. Li, Q. Lu, Q. Wang, J. C. Yu, and Z. Zeng, Angew. Chem. Int. Ed. 62, e202218016 (2023).

\bibitem{ref6}Q. Yun, L. Li, Z. Hu, Q. Lu, B. Chen, and H. Zhang, Adv. Mater. 32, 1903826 (2020).



\bibitem{ref9} D. Lembke, S. Bertolazzi, and A. Kis, Acc. Chem. Res. 48, 100 (2015).
\bibitem{ref10} C. H. Gong, Y. X. Zhang, W. Chen, J. W. Chu, T. Y. Lei, J. R. Pu, L. P. Dai, C. Y. Wu, Y. H. Cheng, T. Y. Zhai, L. Li, and J. Xiong, Adv. Sci. 4, 1700231 (2017).

\bibitem{ref110} K. F. Mak, J. Shan, and D. C. Ralph, Nat. Rev. Phys. 1, 646 (2019).

\bibitem{ref111} K. Premasiri and X. P. Gao, J.
Phys.: Condens. Matter 31, 193001 (2019).

\bibitem{ref11} A. Pospischil and T. Mueller, Appl. Sci. 6, 78 (2016).

\bibitem{ref12}W. Choi, N. Choudhary, G. H. Han, J. Park, D. Akinwande, and Y. H. Lee, Mater. Today 20, 116 (2017).

\bibitem{ref13} E.C. Ahn, npj 2D Mater. Appl. 4, 17 (2020).

\bibitem{ref14} Y. P. Feng, G. Zhang, Z. Wang, J. Kang, B. Peng, S. Wang, J. Wei, and W. Han, WIREs Comput. Mol. Sci. 7, e1313 (2017).

\bibitem{ref15} X. Li and X. Wu, WIREs Comput. Mol. Sci. 6, 441 (2016).

\bibitem{ref16} A. K. Geim and K. S. Novoselov, Nat. Mater. 6, 183 (2007).

\bibitem{ref17} A. K. Geim, Science 324, 1530 (2009).
\bibitem{ref18}D. R. Cooper, B. D’Anjou, N. Ghattamaneni, B. Harack, I. Hilke, A. Horth, N. Majlis, M. Massicotte, L. Vandsburger, E. Whiteway, and V. Yu, ISRN Nanotechnol. 2012, 501686 (2012).

\bibitem{ref161} Y. Fattasse, M. Mekkaoui, A. Jellal, and A. Bahaoui, Physica E 134, 114924 (2021).

\bibitem{ref162} Y. Fattasse, M. Mekkaoui, A. Jellal, and A. Bahaoui, Eur. Phys. J. B, 95,
128 (2022).

\bibitem{ref163} M. Mekkaoui, Y. Fattasse, and A. Jellal, Physics Letters A 439, 128136 (2022).

\bibitem{ref164} Y. Fattasse, M. Mekkaoui, A. Jellal, and A. Bahaoui, Physica E 148,115634 (2022).

\bibitem{ref165} A. Jellal, R. EL Aitouni, P. Díaz, and D. Laroze, Phys. Scr. 100, 045927 (2025).

	


\bibitem{ref22} A. Kormányos, G. Burkard, M. Gmitra, J. Fabian, and V. Zólyomi, 2D Mater. 2, 049501 (2015).


\bibitem{Chernikov2015}
A.~Chernikov, T.~C. Berkelbach, H.~M. Hill, A.~Rigosi, Y.~Li, O.~B.
Aslan, D.~R. Reichman, M.~S. Hybertsen, and T.~F. Heinz,
Phys. Rev. Lett. {113}, 076802 (2014).

\bibitem{Wang2018}
G.~Wang, A.~Chernikov, M.~M. Glazov, T.~F. Heinz, X.~Marie, and B.~Urbaszek,
Rev. Mod. Phys. {90}, 021001 (2018).

\bibitem{Buttiker1983}
M.~B\"uttiker,
Phys. Rev. B {27}, 6178 (1983).

\bibitem{Hauge1989}
E.~H. Hauge and J.~A. St\o vneng,
Rev. Mod. Phys. {61}, 917 (1989).

%
%




 \bibitem{sil}Y. Fattasse, H. Bahlouli, C. Cortes, D. Laroze, and A. Jellal. Ann. Phys., 538, e70194 (2026).
	
	\bibitem{grap}Y. Ban, L.-J. Wang, and X. Chen. J. Appl. Phys., 117, 164307 (2015).



\bibitem{ref23} M. I. Katsnelson, K. S. Novoselov, and A. K. Geim, Nat. Phys. 2, 620 (2006).

\bibitem{ref24}
Y. Li, J. Ludwig, T. Low, A. Chernikov, X. Cui, G. Arefe, Y. D. Kim, A. M. van der Zande, A. Rigosi, H. M. Hill, S. H. Kim, J. Hone, Z. Li, D. Smirnov, and T. F. Heinz, Nat. Phys. 11, 148 (2015).

\bibitem{ref25}
W.-T. Hsu, Z.-A. Zhao, L.-J. Li, C.-H. Chen, 
M.-H. Chiu, P.-S. Chang, Y.-C. Chou, and W.-H. Chang, Nat. Commun. 6, 8963 (2015).

\bibitem{ref26}
T. Yan, X. Qiao, X. Liu, P. Tan, and X. Zhang, 
Sci. Rep. 5, 15625 (2015).

\bibitem{ref27}
S. Zheng, Y. Li, X. Zhang, and H. Zhao, 
Nano Res. 16, 12345 (2023).

\bibitem{ref28}
J. Zhang, K. Wang, Y. Liu, and F. Chen, 
Phys. Rev. B 109, 045401 (2024).

\bibitem{exp1}
J. D. Lu, H. Y. Liu, and S. J. Peng, J. Magn. Magn. Mater. 489, 165478 (2019).

\bibitem{exp2}
J. D. Lu, B. Xu, Y. B. Li, H. Y. Liu, J. Li, and W. Zheng. Vacuum 96, 22 (2013).
\bibitem{WSe2valeurs}
M. Tahir, P. M. Krstajić, and P. Vasilopoulos, Phys. Rev. B 95, 235402 (2017).

\bibitem{Zemman}
H. Sakai, H. Fujimura, S. Sakuragi, M. Ochi, R. Kurihara, A. Miyake, M. Tokunaga, T. Kojima, D. Hashizume, and T. Muro, Phys. Rev. B 101, 081104 (2020).

\bibitem{gs}
G. Aivazian, Z. Gong, A. M. Jones, R.-L. Chu, J. Yan, D. G. Mandrus, C. Zhang, D. Cobden, W. Yao, and X. Xu, Nat. Phys. 11, 148 (2015).
\bibitem{gv}
A. Srivastava, M. Sidler, A. V. Allain, D. S. Lembke, A. Kis, and A. Imamoglu, Nat. Phys. 11, 141 (2015).

\bibitem{infini1}
M. Barbier, M. J. Pereira Jr., P. Vasilopoulos, and F. M. Peeters, Phys. Rev. B 77, 115446 (2008).

\bibitem{infini2}
M. Ramezani Masir, P. Vasilopoulos, and F. M. Peeters, J. Phys.: Condens. Matter 22, 465302 (2010).

\bibitem{chen08} X. Chen, C.-F. Li, and Y. Ban, Eur. Phys. J. B 62, 453
(2008).

\bibitem{Beenakker} C. W. J. Beenakker, R. A. Sepkhanov, A. R. Akhmerov, and J. Tworzydlo, Phys. Rev. Lett. 102, 146804 (2009).




\bibitem{Steinberg11}A. M. Steinberg and R. Y. Chiao, Phys. Rev. A 49, 3283 (1994).
\bibitem{Li111}C.-F. Li, Phys. Rev. A 65, 066101 (2002).

\bibitem{ref19}
C. W. J. Beenakker, Rev. Mod. Phys. 80, 1337 (2008).

\bibitem{Song2012}
Y. Song, H.-C. Wu, and Y. Guo,
Appl. Phys. Lett. 100, 253116 (2012).


\bibitem{Novoselov2005}
K. S. Novoselov, A. K. Geim, S. V. Morozov, D. Jiang, Y. Zhang, S. V. Dubonos, I. V. Grigorieva, and A. A. Firsov, Nature 438, 197 (2005).

\bibitem{CastroNeto2009}
A. H. Castro Neto, F. Guinea, N. M. R. Peres, K. S. Novoselov, and A. K. Geim, Rev. Mod. Phys. 81, 109 (2009).

\bibitem{Kane2005}
C. L. Kane and E. J. Mele,
Phys. Rev. Lett. 95, 226801 (2005).

\bibitem{Song2013}
Y. Song and H.-C. Wu,
J. Phys.: Condens. Matter 25, 355301 (2013).

\bibitem{Wu2013}
W. Wu, W. Zhai, and Y. Yu, Superlattices Microstruct. 60, 240 (2013).


\bibitem{Xiao2012}
D. Xiao, G.-B. Liu, W. Feng, X. Xu, and W. Yao,
Phys. Rev. Lett. 108, 196802 (2012).

\bibitem{Mak2014}
K. F. Mak, K. He, J. Shan, and T. F. Heinz,
Nat. Nanotechnol. 7, 494 (2012).

\bibitem{Xu2014}
X. Xu, W. Yao, D. Xiao, and T. F. Heinz,
Nat. Phys. 10, 343 (2014).

\bibitem{Zibouche2014}
N. Zibouche, A. Kuc, and T. Heine,
Eur. Phys. J. B 85, 49 (2012).



\bibitem{Avsar2014}
A. Avsar, T. Taniguchi, K. Watanabe, R. Nair, G. F. Schneider, A. H. Castro Neto, B. Özyilmaz, and A. C. Ferrari, Nat. Commun. 5, 4875 (2014).

\bibitem{Schaibley2016}
J. R. Schaibley, H. Yu, G. Clark, P. Rivera, A. J. Ross, K. L. Seyler, W. Yao, and X. Xu, Nat. Rev. Mater. 1, 16055 (2016).

\bibitem{wu2009Conc}X. Chen, J.-W, Tao1, and Y. Ban Eur. Phys. J. B 79, 203–208 (2011).

\bibitem{chen2011Conc} Z. Wu, K. Chang, J. T. Liu, and X. J. Li, J. Appl. Phys. 105, 043702 (2009).

\bibitem{Yes2016}
C. Yesilyurt and S. G. Tan, AIP Adv. 6, 056303 (2016).

\end{thebibliography}
\end{document}